\documentclass[12pt,a4paper]{article}
\usepackage{graphicx,color,amsmath, amssymb,cite,multirow}
\usepackage[english]{babel}
\usepackage{t1enc}
\usepackage[latin2]{inputenc}
\begin{document}
\textwidth=135mm
 \textheight=200mm

\begin{center}
{\bfseries A method of $\eta'$ decay product selection to detect partial chiral symmetry restoration
\footnote{{\small Talk at the VI Workshop on Particle Correlations and Femtoscopy, Kiev, September 14-18, 2010.}}}
\vskip 5mm
M. Csanád$^\dag$ M. Kőfaragó$^\dag$
\vskip 5mm
{\small {\it $^\dag$ Eötvös University, Department of Atomic Physics, Pázmány Péter s. 1/A, H-1117 Budapest, Hungary}}
\\
\end{center}
\vskip 5mm
\centerline{\bf Abstract}

In case of chiral U$_A(1)$
symmetry restoration the mass of the $\eta'$ boson (the ninth, would-be
Goldstone boson) is decreased, thus its production cross section is
heavily enhanced~\cite{Kapusta:1995ww}. The $\eta'$ decays (through one of its decay channels)
into five pions. These pions will not be correlated in terms of Bose-Einsten correlations,
thus the production enhancement changes the strength of two-pion
correlation functions at low momentum~\cite{Vance:1998wd}. Preliminary
results strongly support the mass decrease of the $\eta'$ 
boson~\cite{Csanad:2005nr,Vertesi:2009ca,Csorgo:2009pa}.
In this paper we propose a method to select pions coming from $\eta'$ decays.
We investigate the efficiency of the proposed kinematical cut in several
collision systems and energies with several simulators.
We prove that our method can be used in all investigeted collision systems.
\vskip 10mm

\section{Chiral symmetry and $\eta'$ mass}
The temperature of the quark gluon plasma created in gold-gold collisions of the Relativistic Heavy Ion Collider (RHIC)
may reach values up to 300-600 MeV~\cite{Adcox:2004mh,Adare:2008fqa}. At these very high temperatures the degrees of freedom
are not hadrons but quarks or gluons. It is expected that the broken symmetries of QCD are partially restored in this matter.
Originally, the $U_A(1)$ part of chiral symmetry is broken exactly, thus a high mass meson is produced, the $\eta'$
particle, which has a mass of 958 MeV. However, if the $U_A(1)$ symmetry is partially
restored, the mass of the $\eta'$ is decreasing~\cite{Kapusta:1995ww}.

Production cross sections of hadrons are exponentially suppressed by their mass. Hence without mass modification roughly two orders of
magnitute less $\eta'$ mesons are produced than pions. In contrast, decreased mass $\eta'$ mesons will be created more abundantly.
Thus the number of $\eta'$ mesons is closely related to their mass.

The decay of the $\eta'$ happens after it regained its vacuum mass (at the expense of its momentum).
One important decay channel is the decay into two leptons, $\eta'\rightarrow{l^++l^-}$,
this is investigated in ref.~\cite{Adare:2009qk}. It turns out that there is an excess in the dilepton spectrum at low invariant mass, and this excess
might be related to the $\eta'$ enhancement. There is also a decay mode when the $\eta'$ goes into an $\eta$ and two pions, and the $\eta$ also decays
into three pions:
\begin{align}
\eta'\rightarrow\eta+\pi^++\pi^-\rightarrow\left(\pi^++\pi^-+\pi^0\right)+\pi^++\pi^-\label{e:decay}
\end{align}
and the overall probability of this decay chain is 10\%~\cite{Nakamura:2010zzi}. The average momentum of the resulting five pions is $138$ MeV
due to the low momentum of the original $\eta'$~\cite{Vance:1998wd}.

\section{Two-pion Bose-Einstein correlations}
Final state effects distort two-particle correlation functions. One of the most important final state effect is that of Bose-Einstein correlations.
To calculate these correlation functions, let us utilize the core-halo model.
In the core-halo model~\cite{Csorgo:1999sj}, the hadronic source is divided into two parts: a core and a halo. The core consists of the
primordial particles and decay products of very short lifetime resonances, thus its size is very small, roughly 5 fm. The halo consists
of decay products of long lived resonances, such as $\eta$, $\eta'$ or $K^0_S$. The halo hadrons
are created very far from the core, so the halo size is much larger, at least 50 fm. When measuring correlation
functions however, due to finite momentum resolution of the detectors very small momentum differences cannot be resolved, i.e.
pairs with such similar momenta are regarded as one by the detectors. In the Fourier transformation, large sizes correspond to
small momenta, thus the halo correlations are not seen in measurements. Hence the observable part of the correlation
function is due to the core, but its strength is decreased by the core/halo ratio.
In case of a plain-wave approximation, for an identical boson pair with momenta $p_1$ and $p_2$,
the two-particle correlation function is (for details see e.g. ref.~\cite{Csorgo:1999sj}):
\begin{align}
C_2(q,K)=1+\lambda\frac{{\vert\widetilde{S}_C(q,K)\vert}^2}{{\vert\widetilde{S}_C(q=0,K)\vert}^2}
\end{align}
where $q=p_1-p_2$, $K=(p_1+p_2)/2$ and $\widetilde{S_C}(q,p)$ is the Fourier transformed of the core source function (or emission function
of pions coming from the core) $S_C(x,K)$ (the Fourier transformation is in $x\rightarrow q$). The important parameter here is
$\lambda=\left.N_C\middle/(N_C+N_H)\right.$, where the number of particles in the core is $N_C$, number of particles in the halo is $N_H$.
The $\lambda$ parameter thus depends on the ratio of the core to the halo. The halo pions come partly from $\eta'$ decays, so
more $\eta'$ means larger halo, and smaller $\lambda$. Hence $\eta'$ mass and $\lambda$ value are connected.
This was investigated in refs.~\cite{Csanad:2005nr,Csorgo:2009pa,Vertesi:2009ca}, and it was found that the $\lambda$ parameter is indeed
decreasing at the kinematical domain of $\eta'$ decay pions. However, it is not experimentally proven that
the $\eta'$ decay pions are causing the decrease. In this paper we investigate a method to kinematically filter out pions from
$\eta'$ decays. If applied to the experimental sample, in case of an $\eta'$ mass modification the $\lambda$ decrese will vanish.

\section{Kinematical domain of pions from $\eta'$ decays}

Our method is based on the invariant mass of pions from the decay chain mentioned in eq.~\ref{e:decay}.
Let us investigate the invariant momentum of $\pi^+$,$\pi^-$ pairs in the above decays. Using the mass-shell
condition for pions, and utilize momentum conservation in the decay, one gets an $m_{inv}$ interval for the $\pi^+$,$\pi^-$ pair coming
from the $\eta'$, for the second pion pair coming from the $\eta$, and for the whole quadruplet. Based on the simulations, we chose the
$0.075$--$0.171$ GeV$^2/c^4$ interval for pairs and the $0.43$--$0.69$ GeV$^2/c^4$ interval for
quadruplets. These yields an effective method of kinematical selection of $\eta'$ decay pions, as detailed in the next section.

Our method is the following. For a given $\pi^+$, we take all possible pion quadruplets ($\pi^+$,$\pi^-$,$\pi^+$,$\pi^-$) from the same event, and check
if any of these quadruplets fulfill our $m_{inv}$ criteria. If there is at least one, we consider the original $\pi^+$ to be ``found'', or say
that if fulfills our $m_{inv}$ criteria. The same can be done by starting from a given $\pi^+$,$\pi^-$ pair. In this case ``finding'' a non-$\eta'$
pion pair will be less probable, but the efficiency of finding $\eta'$ pions will not be different. So this is expected to be a better working method.
In simulations, we can determine if the pair of the particle comes from an $\eta'$ decay, so the efficiency of the selection method
can be tested. Whether using the pair or the single particle method, we can form four different groups of them:
\begin{itemize}
\addtolength{\itemsep}{-0.5\baselineskip}
\item[a)] Comes from an $\eta'$ and fulfills the $m_{inv}$ criteria
\item[b)] Comes from an $\eta'$ and does not fulfill the $m_{inv}$ criteria
\item[c)] Does not come from an $\eta'$ and fulfills the $m_{inv}$ criteria
\item[d)] Does not come from an $\eta'$ and does not fulfill the $m_{inv}$ criteria
\end{itemize}
Ideally all $\eta'$ pions go into the first group, while all others fall into the last group. Let us call the number of pions (or pairs in the other method)
in the four groups $N_a$, $N_b$, $N_c$ and $N_d$, respectively. In the next part, we will give the following numbers for each type of simulation:
Efficiency of cutting $\eta'$ decay products, $\left.N_a\middle/(N_a+N_b)\right.$; and loss of statistics, i.e. fraction of lost non-$\eta'$ pions, 
$\left.N_c\middle/(N_c+N_d)\right.$. The optimal value for efficiency is 1 (in this case we could cut out all pions coming from $\eta'$ mesons),
0 for loss (in this case we kept all non-$\eta'$ pions). Note that a ``found'' pair or particle that is not coming from
an $\eta'$ (i.e. a loss greater than zero) decreases our experimental sample. If the loss is 90\%, then the sample is reduced by a factor of 10,
so the statistical errors will heavily increase. Goal of present paper is to investigate the efficiency and loss connected to our the method.
A similar method was investigated in ref.~\cite{Kulka:1990zh} for $e^+ e^-$ collisions. We test the method in p+p and Au+Au collisions, at several center-of-mass
energies.

\begin{figure}
\begin{center}
\includegraphics[width=0.71\linewidth]{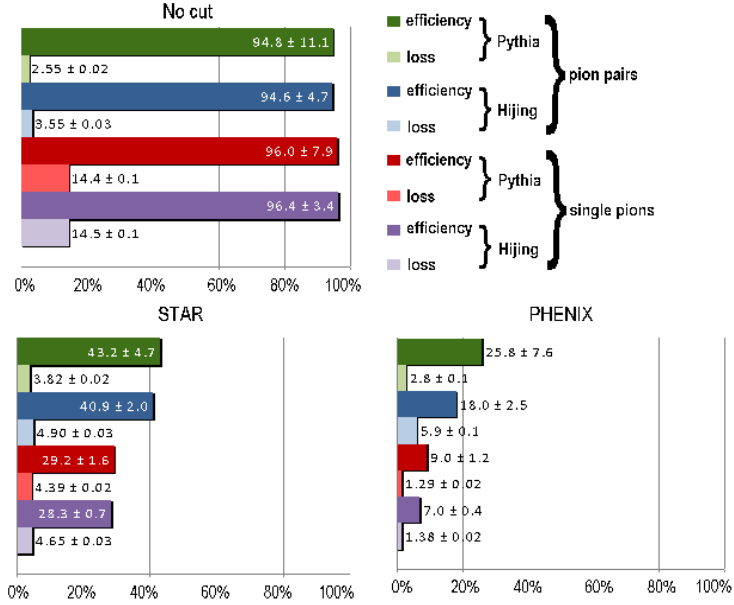}
\end{center}
\caption{Results from $\sqrt{s}=200$ GeV p+p collisions. The method is working in all cases, but efficiency is very low in case of PHENIX cuts,
         for the single particle method.}
\label{f:200pp}
\end{figure}

\begin{figure}
\begin{center}
\includegraphics[width=0.71\linewidth]{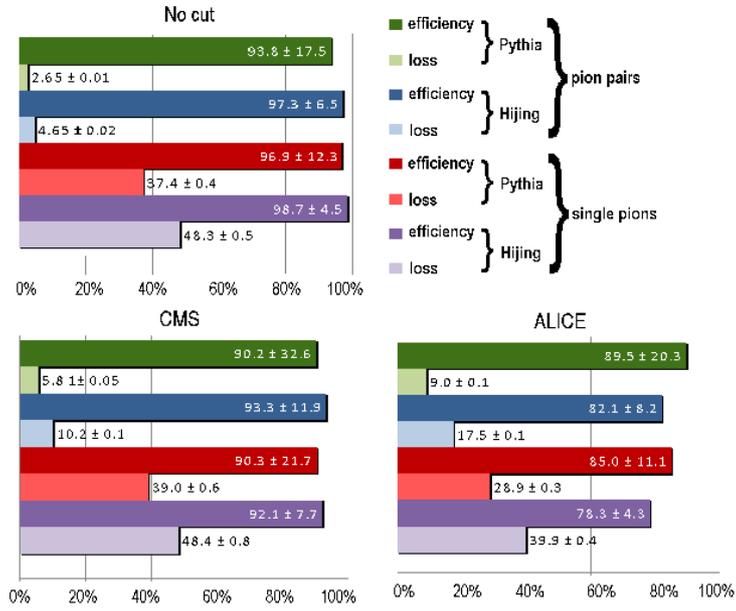}
\end{center}
\caption{Results from $\sqrt{s}=200$ GeV p+p collisions. The method is working in all cases.}
\label{f:14000pp}
\end{figure}

\begin{figure}
\begin{center}
\includegraphics[width=0.71\linewidth]{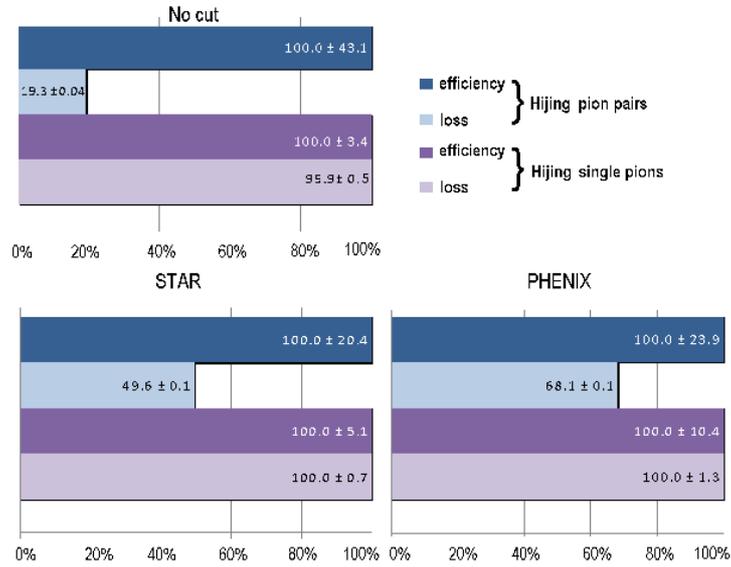}
\end{center}
\caption{Results from $\sqrt{s}=200$ GeV Au+Au collisions. The method is working only in the pair method, as loss is 100\% in the other case.}
\label{f:200auau}
\end{figure}

\section{Results}

We used two simulations to test our method: Pythia~\cite{Sjostrand:2007gs} (proton-proton collisions, version 8.135) and
Hijing~\cite{Gyulassy:1994ew} (proton-proton and gold-gold collisions, version 1.411). We also simulated the geometric
acceptance of the detectors. In case of the 200 GeV RHIC energy, we used the geometry of STAR and PHENIX detectors, while in case of
14 TeV energy, we used the geometry of ALICE and CMS detectors.

We generated 1 000 000 $\sqrt{s}=200$ GeV and 10 000 $\sqrt{s}=14$ TeV p+p events. For PHENIX cuts, the efficiency is the lowest,
because a large fraction of pions are not detected. At both energies, the pair method is better than the single particle method.
We also generated 100 $\sqrt{s_{NN}}=200$ GeV events. Here only the pair method is working, as
essentially all single particles are found, due to the very large statistics. See details on figs.~\ref{f:200pp}-\ref{f:200auau} for details.
\bibliographystyle{prlstyl}
\bibliography{../../../../master}

\begin{thebibliography}{10}

\bibitem{Kapusta:1995ww}
J.~I. Kapusta {\it et~al.}, Phys. Rev. {\bf D53},  5028  (1996)

\bibitem{Vance:1998wd}
S.~E. Vance {\it et~al.}, Phys. Rev. Lett. {\bf 81},  2205  (1998)

\bibitem{Csanad:2005nr}
M. Csan\'ad, Nucl. Phys. {\bf A774},  611  (2006)

\bibitem{Vertesi:2009ca}
R. V\'ertesi {\it et~al.}, Nucl.Phys. {\bf A830},  631C  (2009)

\bibitem{Csorgo:2009pa}
T. Cs\"org\H{o} {\it et~al.}, Phys.Rev.Lett. {\bf 105},  182301  (2010)

\bibitem{Adcox:2004mh}
K. Adcox {\it et~al.}, Nucl. Phys. {\bf A757},  184  (2005)

\bibitem{Adare:2008fqa}
A. Adare {\it et~al.}, Phys. Rev. Lett. {\bf 104},  132301  (2010)

\bibitem{Adare:2009qk}
A. Adare {\it et~al.}, Phys. Rev. {\bf C81},  034911  (2010)

\bibitem{Nakamura:2010zzi}
K. Nakamura {\it et~al.}, J. Phys. {\bf G37},  075021  (2010).

\bibitem{Csorgo:1999sj}
T. Cs\"org\H{o}, Heavy Ion Phys. {\bf 15},  1  (2002)

\bibitem{Kulka:1990zh}
K. Kulka and B. Lorstad, Nucl. Instrum. Meth. {\bf A295},  443  (1990).

\bibitem{Sjostrand:2007gs}
T. Sjostrand {\it et~al.}, Comput.Phys.Commun. {\bf 178},  852  (2008)

\bibitem{Gyulassy:1994ew}
M. Gyulassy and X.-N. Wang, Comput. Phys. Commun. {\bf 83},  307  (1994)

\end{thebibliography}

\end{document}